\documentclass[journal=cmatex,layout=draft]{achemso}
\setcitestyle{square}
\usepackage[version=4]{mhchem}
\usepackage[hyperindex]{hyperref}
\hypersetup{colorlinks,citecolor=blue,linkcolor=blue,urlcolor=blue}
\usepackage{units}
\usepackage{amsmath}
\usepackage{booktabs}
\usepackage{etoolbox}

\newcommand{\diffd}{\,\text{d}}
\newcommand{\voc}{V_\text{OC}}
\newcommand{\vpre}{V_\text{pre}}
\newcommand{\ndark}{n_\text{dark}}
\newcommand{\kd}{k_d}
\newcommand{\kenc}{k_\text{enc}}
\newcommand{\kexc}{k_\text{exc}}
\newcommand{\kct}{k_\text{CT}}

\makeatletter
\patchcmd{\acs@contact@details}{E}{*\,E}{}{}
\makeatother

\author{Sebastian Wilken}
\affiliation{Department of Physics, Faculty of Science and Engineering, \AA{}bo Akademi University, Porthansgatan 3, 20500 Turku, Finland}
\alsoaffiliation{Department of Physics, Chemistry and Biology, Link\"oping University, 581 83 Link\"oping, Sweden}
\email{sebastian.wilken@abo.fi}

\author{Tanvi Upreti}
\affiliation{Department of Physics, Chemistry and Biology, Link\"oping University, 581 83 Link\"oping, Sweden}

\author{Armantas Melianas}
\affiliation{Department of Materials Science and Engineering, Stanford University, Stanford, CA 94305, USA}

\author{Staffan Dahlstr\"om}
\affiliation{Department of Physics, Faculty of Science and Engineering, \AA{}bo Akademi University, Porthansgatan 3, 20500 Turku, Finland}

\author{Gustav Persson}
\author{Eva Olsson}
\affiliation{Department of Physics, Chalmers University of Technology, 412 96 G\"oteborg, Sweden}

\author{Ronald \"{O}sterbacka}
\affiliation{Department of Physics, Faculty of Science and Engineering, \AA{}bo Akademi University, Porthansgatan 3, 20500 Turku, Finland}

\author{Martijn Kemerink}
\affiliation{Department of Physics, Chemistry and Biology, Link\"oping University, 581 83 Link\"oping, Sweden}

\title{Experimentally Calibrated Kinetic Monte Carlo Model Reproduces Organic Solar Cell Current--Voltage Curve}

\SectionNumbersOn

\begin{document}

\begin{abstract}
Kinetic Monte Carlo~(KMC) simulations are a powerful tool to study the dynamics of charge carriers in organic photovoltaics. However, the key characteristic of any photovoltaic device, its current--voltage~($J$--$V$) curve under solar illumination, has proven challenging to simulate using KMC. The main challenges arise from the presence of injecting contacts and the importance of charge recombination when the internal electric field is low, i.e., close to open-circuit conditions. In this work, an experimentally calibrated KMC model is presented that can fully predict the $J$--$V$ curve of a disordered organic solar cell. It is shown that it is crucial to make experimentally justified assumptions on the injection barriers, the blend morphology, and the kinetics of the charge transfer state involved in geminate and nongeminate recombination. All of these properties are independently calibrated using charge extraction, electron microscopy, and transient absorption measurements, respectively. Clear evidence is provided that the conclusions drawn from microscopic and transient KMC~modeling are indeed relevant for real operating organic solar cell devices.
\end{abstract}

\section{Introduction}
Kinetic Monte Carlo~(KMC) simulations have successfully been used to model the charge carrier dynamics in organic photovoltaics~(OPVs) on the ps to $\mu$s time scale. For instance, it was shown that in thin-film OPV~devices, thermalization in the disorder-broadened density of states~(DOS) does not complete before charges are extracted.\cite{Melianas2014,Howard2014,Melianas2015,Melianas2017} The conclusions from these studies are drawn from the fitting of time-resolved experiments performed under certain bias conditions such as short circuit or open circuit. Other authors used KMC modeling to focus on the process of charge recombination and its dependence on the morphology in slabs of material, i.e., in absence of contacts.\cite{Lyons2012,Heiber2015,Groves2008,Coropceanu2017,Kaiser2019} However, it is still an open question to which extent nonequilibrium phenomena and other aspects that are not accounted for in macroscopic simulations such as quasi-equilibrium drift--diffusion~(DD) models, govern the steady-state operation of complete OPV~devices. To answer the question, it would be highly desirable to have a microscopic model that is also able to describe the current--voltage~($J$--$V$) curve, particularly the open-circuit voltage~($\voc$) and the fill factor.

Modeling $J$--$V$~curves with KMC has so far proven nearly impossible. One of the main challenges is the presence of two injecting contacts. While it may be acceptable to consider the contacts as simple sinks for electrons and holes in transient extraction experiments~(performed at $V \ll \voc$), this simplification does not work for situations closer to~$\voc$. When the internal field is low, contacts inject many charge carriers into the active layer. This high carrier density is demanding from the computational point of view and challenging to correctly account for. Even though a few concepts exist how contacts can be implemented in KMC, literature studies have so far failed to fully describe~$J$--$V$ data of real devices or are based on assumptions that are not justified experimentally.\cite{Meng2010,Meng2011,Kipp2013} 

Besides computational challenges, the injected charge density also sets the boundary conditions for the recombination of photogenerated carriers.\cite{Wurfel2019} Charge recombination generally becomes more important when going from short circuit to open circuit because transport will slow down. Indeed, the competition between charge extraction and recombination has been demonstrated to be the main determinant of the device fill factor.\cite{Neher2016,Bartesaghi2015,Kaienburg2016} For a device model to be reliable it must therefore capture the hopping transport characteristics and the recombination kinetics at the same time. Even though the mechanisms of charge recombination are highly disputed, it is commonly accepted that the morphology plays a key role.\cite{Gohler2018,Lakhwani2014,Heiber2015} For instance, it is well documented that aggregated donor or acceptor domains may lower the recombination rate.\cite{Burke2014,Jamieson2012,Sweetnam2014,McMahon2011} However, although the morphology of many donor/acceptor blends is well characterized by electron microscopy and other techniques, the nanostructure is often neglected in KMC and an effective medium is assumed instead.\cite{Melianas2019}

Here, we present a KMC model that successfully predicts device $J$--$V$ curves while simultaneously accounting for nonequilibrium hopping transport and recombination dynamics. We show that this is only possible when correct assumptions are made on the injection barriers, the morphology of the active layer, and the charge recombination rate. All these properties are calibrated by independent experimental techniques such as charge extraction, electron microscopy and transient absorption. We are thereby introducing a device model that works on a multitude of length and time scales. As such it will be useful for future investigations on the interplay between elementary processes and device characteristics of organic solar cells and other optoelectronic devices.

\section{Results and Discussion}
\subsection{Material System}
\begin{figure*}[h!]
\includegraphics{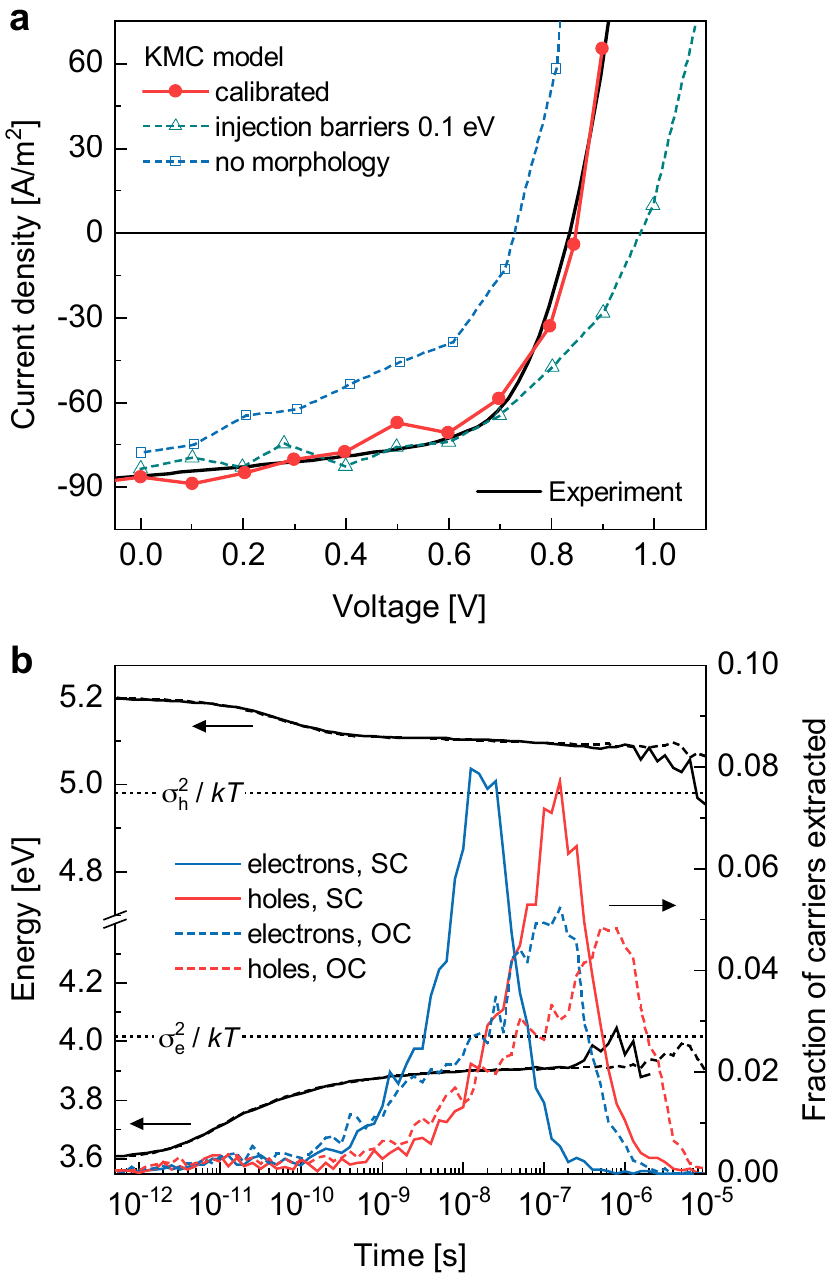}
\caption{Key results of the KMC model described in this work. (a)~Experimental and simulated $J$--$V$ curves of a TQ1:P\ce{C71}BM solar cell~(active-layer thickness:~\unit[72]{nm}) with the device architecture~ITO\slash{}PEDOT:PSS\slash{}TQ1:P\ce{C71}BM\slash{}LiF\slash{}Al under simulated sunlight. Filled circles refer to the calibrated KMC model with the input parameters given in Table~\ref{tab:params}. Open symbols are simulations with the same parameter set, but assuming too low injection barriers of \unit[0.1]{eV}~(triangles) or only an effective medium without P\ce{C71}BM aggregates~(squares). (b)~Simulated relaxation~(black lines, left axis) and extraction time distributions~(colored lines, right axis) of photogenerated charges under steady-state illumination at short circuit~(SC, solid lines) and open circuit~(OC, dashed lines) using the experimentally calibrated KMC~model. Both electrons and holes are extracted before they reach their quasi-equilibrium energy~(dotted lines).}
\label{fig:figure1}
\end{figure*}

The aim of this work is to develop and experimentally calibrate a KMC~model that fits both transient experiments and device $J$--$V$ curves. Our material system for experimental calibration is~TQ1:P\ce{C71}BM,\cite{Wang2010} an archetypal polymer/fullerene blend. The reason for choosing~TQ1:P\ce{C71}BM is that for this specific system a clear picture of the carrier dynamics has emerged from time-resolved measurements and previous modeling, which is summarized in a recent review article.\cite{Melianas2019} Hence, many of the parameters for the KMC model are already known; in particular, it has been shown that the charge extraction in thin devices with an active-layer thickness~$\approx\unit[100]{nm}$ is strongly affected by nonequilibrium effects. Figure~\ref{fig:figure1} shows the experimental $J$--$V$~curve of a 72-nm thick TQ1:P\ce{C71}BM solar cell under simulated sunlight. The device displays an open-circuit voltage of~\unit[835]{mV}, a short-circuit current of~$\unit[86]{A\,m^{-2}}$, a fill factor of~0.63, and an efficiency of~4.5\%.

\subsection{KMC Describes Device Current--Voltage Curve}
The KMC model, which is extended and experimentally calibrated in this work to fully describe OPV devices, has been introduced previously.\cite{Melianas2014,Melianas2015} Briefly, it implements the extended Gaussian disorder model on a simple cubic lattice and takes into account: excitons; charge transfer~(CT) pairs; electrons and holes; morphology via the allocation of individual hopping sites to different material phases; charge injection/extraction by hopping from/to the Fermi level of the respective contact; full Coulomb interactions, including those by image charges in the electrodes; periodic boundary conditions in the lateral directions.

Charge transport is described in terms of the Miller--Abrahams model, in which the hopping rate~$\nu_{ij}$ from site~$i$ to site~$j$ separated by a distance~$r_{ij}$ is given by
\begin{equation}
\nu_{ij} = \nu_0\exp(-2\alpha r_{ij}) \begin{cases}
    \exp\left(-\frac{\Delta E_{ij}}{kT}\right)& \Delta E_{ij} > 0\\
    1              & \Delta E_{ij} \leq 0,
\end{cases}
\label{eq:eq1}
\end{equation}
where $\nu_0$ is the attempt-to-hop frequency, $\alpha$ the inverse localization length, $\Delta E_{ij} = E_j - E_i$ the energy difference between the sites, and $kT$ the thermal energy. Hopping is assumed to take place in a Gaussian~DOS,
\begin{equation}
g(E) = \frac{1}{\sqrt{2\pi\sigma^2}}\exp\left[-\frac{(E - E_0)^2}{2\sigma^2}\right],
\end{equation}
where~$E$ is the single particle energy, $E_0$ the mean energy, and $\sigma$ the width of the Gaussian~DOS or the energetic disorder. We note that without of loss of generality, also other energy distributions could be assumed in the model, such as an exponential~DOS. From previous studies, however, it is known that a Gaussian~DOS gives the most appropriate description for the present TQ1:P\ce{C71}BM~system both when describing transient and steady-state experiments.\cite{Melianas2019,Melianas2019b} In this work, only hopping between nearest neighbors on a regular, six-fold coordinated lattice was considered. In this configuration, the localization length~$\alpha$ is unimportant; the first exponential term of Equation~(\ref{eq:eq1}) was implicitly included in $\nu_0$, that is, the rate of downward nearest-neighbor hops.

The core working principle of a KMC model is to simulate the time evolution of a system based on the transition rates of all possible events~(here: hops, generation, recombination, injection). The event that occurs at a certain point in time is randomly selected with the transition rates used as weighting factors. The time step between single events is calculated as $\tau = -\ln(u)/\Sigma_\nu$, where $u$ is a random number drawn from a homogeneous distribution between 0 and 1, and $\Sigma_\nu$ the sum of the rates of all possible events. A typical simulation starts with a number of photogenerated excitons. Excitons may separate into CT pairs/free charges or recombine after their lifetime. Diffusion of excitons by F\"orster and Dexter energy transfer are both accounted for. An additional on-site barrier of~\unit[0.8]{eV} is used to facilitate charge separation in molecularly mixed phases. Further details on the used KMC algorithm can be found in the Supplementary Information of Ref.~\citenum{Melianas2015}.

As mentioned above, the presence of injecting contacts causes computational challenges. Charge injection is mediated by the injection barriers, i.e., the energy offset between the contact Fermi level and the respective molecular orbital of the semiconductor. Especially for low barriers, carriers may oscillate multiple times across the contact interface before injection/extraction finally takes place. We mitigated this `small barrier' problem by only allowing for a transfer if the number of charges next to the contact interface deviates from its equilibrium value. The transfer is modeled as hopping event with an attempt frequency~$\nu_{0,\text{cont}}$ of the same order as for the transport of the faster carrier~(here: electrons) in the semiconductor. This ensures that charge collection is not limited by the contacts. Both the cathode and anode were considered nonselective; hence, possible losses due to diffusion of carriers into the `wrong' contact are implicitly accounted for.

An advantage of KMC simulations is that no explicit assumptions about the formalism of charge recombination need to be made. Recombination of free charges involves the formation of a CT~pair as intermediate. Exciton formation is explicitly allowed, but requires overcoming the relevant energy level offset between the TQ1 and P\ce{C71}BM; as such, it can be interpreted as the inverse of charge separation, i.e., the splitting of~(CT)~excitons into free electrons and holes. As discussed in more detail below, it is then the inverse lifetime of the CT state that determines the recombination rate and must be calibrated experimentally.

\begin{table}[t]
\caption{Key parameters used in the calibrated KMC model.}
\begin{tabular}{lc}
\toprule
Parameter & Value\\
\midrule
Simulated volume [sites] & $40 \times 40 \times 40$ \\
Nearest neighbor distance, $a_\text{NN}$ [nm] & $1.8$ \\
Energetic disorder electrons, $\sigma_\text{e}$ [meV] & $75$ \\
Attempt-to-hop frequency electrons, $\nu_{0,\text{e}}$ [$\unit{s^{-1}}$] & $1 \times 10^{11}$ \\
Energetic disorder holes, $\sigma_\text{h}$ [meV] & $75$ \\
Attempt-to-hop frequency holes, $\nu_{0,\text{h}}$ [$\unit{s^{-1}}$] & $1 \times 10^{10}$ \\
Inverse exciton lifetime, $\kexc$ [$\unit{s^{-1}}$]	& $1 \times 10^{9}$ \\
Inverse CT~state lifetime, $\kct$ [$\unit{s^{-1}}$] & $3 \times 10^{7}$ \\
Injection barrier height [eV] &	$0.2$ \\
Contact attempt-to-hop frequency, $\nu_{0,\text{cont}}$ [$\unit{s^{-1}}$] & $1 \times 10^{11}$ \\
\bottomrule
\end{tabular}
\label{tab:params}
\end{table}

The filled circles in Figure~\ref{fig:figure1}a show that after the calibration discussed below, the KMC~model fits the $J$--$V$ curve of the TQ1:P\ce{C71}BM solar cell well within experimental accuracy and matches both the device~$\voc$ and fill factor. Table~\ref{tab:params} lists the key parameters used for the simulations. We note that these values are not the result of a fitting routine but come from independent characterizations. The hopping parameters were chosen in such a way that they represent earlier experiments, such as time-resolved electric-field-induced second harmonic generation~(TREFISH)\cite{Melianas2017} and temperature-dependent space-charge-limited currents~(SCLC),\cite{Melianas2017,Upreti2019} but at the same time allow efficient calculations. This was done by assuming a single disorder for electrons and holes~($\sigma_\text{e} = \sigma_\text{h} \equiv \sigma$) and adjusting the attempt frequencies~$\nu_0$ such that the macroscopic transport characteristics of TQ1:P\ce{C71}BM, e.g., the contrast between electron and hole mobility, are still captured~(see Supporting Information for details). Figure~\ref{fig:figure1}b shows that also with the symmetrized hopping parameters, relaxation in the~DOS is far from being complete when photogenerated carriers are extracted. This is true for both short-circuit and open-circuit conditions, which indicates that nonequilibrium effects may affect charge extraction along the entire $J$--$V$~curve. A detailed discussion of how the nonequilibrium effects influence the individual performance parameters will be the topic of another publication.

The main result of this study is that a KMC~model that can describe full $J$--$V$~characteristics requires an appropriate description and calibration of the injection barriers and the morphology in the active layer. If wrong or too simple assumptions are made on these properties, our otherwise well validated KMC~model can no longer describe the device~(Figure~\ref{fig:figure1}a, open symbols). Because this mainly concerns~$\voc$ and the fill factor, these observations are closely related to the charge recombination. In the following sections we will therefore focus on the factors that determine the shape of the $J$--$V$ curves in the fourth quadrant, that is the injection barrier height, the blend morphology, and the recombination rate. 

\subsection{Calibration of Injection Barriers}
The injection barriers set the carrier density in the device around the built-in voltage. To get a realistic estimate of the barrier height, we compare the results of charge extraction experiments in the dark with device simulations. As KMC~calculations are computationally too expensive for this approach, we used a DD~model instead.\cite{Burgelman2000,Wilken2020} This is justified because the charges treated here were not photogenerated, but injected from the contacts, so that the complexities of exciton/charge separation are bypassed. Furthermore, charges are injected from thermalized reservoirs~(contacts), so that it is reasonable to describe them by a quasi-equilibrium mobility. The mobility values were estimated by inserting the hopping parameters in Table~\ref{tab:params} in the mobility functional by Pasveer~et~al.\cite{Pasveer2005} Charge recombination is assumed to be strictly bimolecular with the steady-state recombination coefficient~($\unit[6 \times 10^{-18}]{m^3\,s^{-1}}$) taken from experimental studies on TQ1:P\ce{C71}BM.\cite{Murthy2013,Roland2019}

\begin{figure*}[t]
\includegraphics{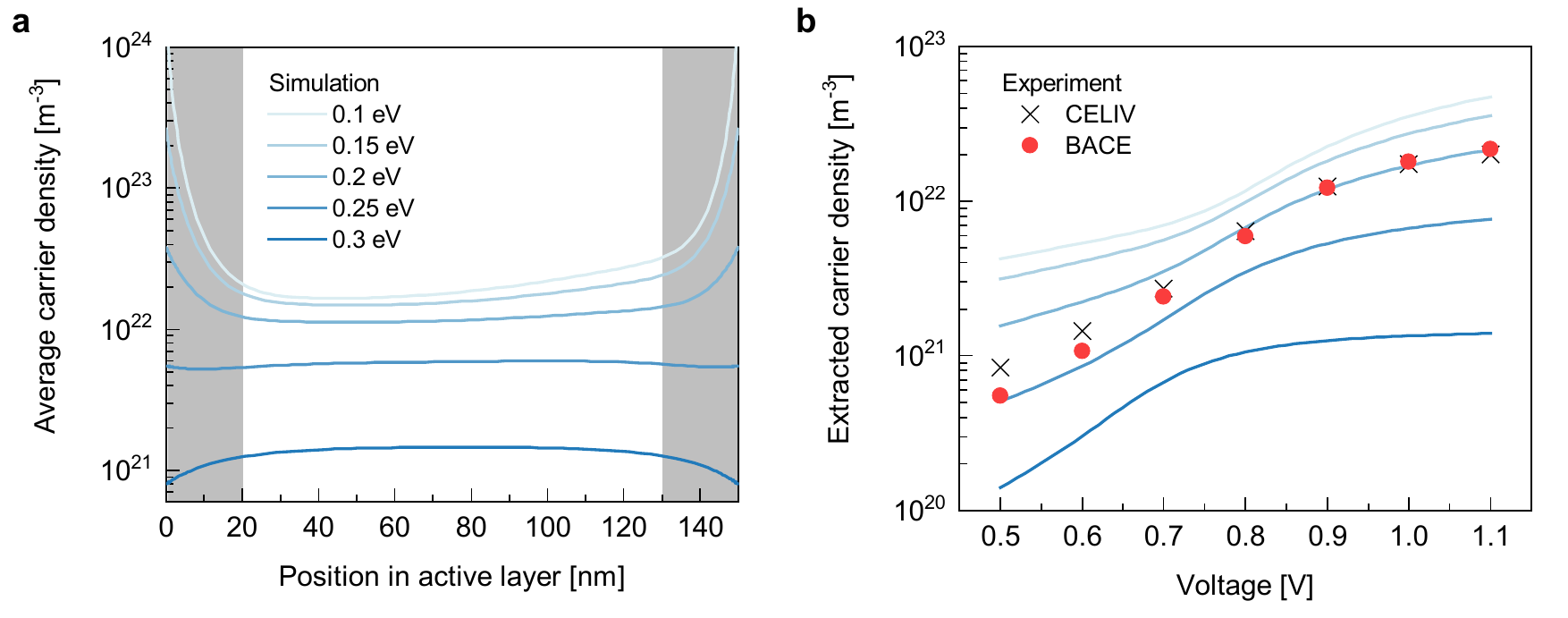}
\caption{Calibration of the injection barrier height. (a)~Spatial carrier density profiles at a forward bias of \unit[0.9]{V} for different barriers at the anode and cathode. Data refers to the average of the electron and hole density calculated with a drift--diffusion model. Gray shaded areas mark the space-charge regions close to the contact interfaces. (b)~Extracted carrier density according to Equation~(\ref{eq:ndark}) from CELIV and BACE experiments on a 150-nm thick TQ1:P\ce{C71}BM solar cell~(symbols). The voltage axis refers to the bias~$\vpre$ present before charges were extracted by the reverse voltage pulse. Colored traces are the prediction from drift--diffusion calculations using the same injection barrier heights as in panel~(a).}
\label{fig:figure2}
\end{figure*}

Figure~\ref{fig:figure2}a illustrates the effect of the injection barrier height on the average carrier density. Here, we chose devices with an active-layer thickness of~\unit[150]{nm}; only at these larger thicknesses a `bulk' region is established, which makes the comparison with charge-extraction experiments more reliable.\cite{Kirchartz2012,Deledalle2014} Note that especially at higher densities the carrier profiles are not perfectly symmetric, which is due to the imbalanced electron and hole transport.\cite{Scheunemann2019,Wilken2020} The experiments to be simulated are charge extraction by linearly increasing voltage~(CELIV) and bias-assisted charge extraction~(BACE). In both techniques, the device is held at a certain pre-bias~($\vpre$) until a steady state is reached; the charges in the device are then extracted by applying a triangular~(CELIV) or rectangular~(BACE) voltage pulse. The dark carrier density is calculated from the transient current~$J(t)$ via
\begin{equation}
\ndark = \frac{1}{qd}\int_0^{t_\text{f}} \left[J(t) - J_0(t)\right] \diffd t,
\label{eq:ndark}
\end{equation}
where $q$ is the elementary charge, $d$ the active-layer thickness, $J_0$ the displacement current measured at $\vpre = 0$, and $t_\text{f}$ the time at which charge extraction is completed. Note that in the form of Equation~(\ref{eq:ndark}), the carrier density represents the average of electrons and holes, as pointed out by Hawks~et~al.\cite{Hawks2015}

Figure~\ref{fig:figure2}b shows that CELIV and BACE give a consistent picture of the carrier density as a function of voltage. At $\vpre = \unit[0.9]{V}$, which approximately corresponds to open-circuit conditions under 1-sun illumination, $\ndark$ is about $\unit[1 \times 10^{22}]{m^{-3}}$. This is the same order of magnitude as for the photogenerated carrier density and indicates the importance of injected carriers for charge recombination. As can be seen, the best description of the dark carrier density and its voltage dependence is obtained for a barrier height of~\unit[0.2]{eV}; with this value, the KMC~model reproduces the experimental $J$--$V$~curve~(see Figure~\ref{fig:figure1}). We note that the discrepancy between CELIV/BACE and DD~simulation at voltages well below the built-in voltage is merely due to experimental limitations. In this regime, most carriers are situated in the thin space-charge regions close to the contacts, which makes them only partly visible to charge-extraction experiments.\cite{Kniepert2014,Neher2016}

If instead too small injection barriers are selected as input for the KMC~model, it can no longer describe both $\voc$ and the fill factor. The open triangles in Figure~\ref{fig:figure1} illustrate this for a barrier height of~\unit[0.1]{eV}. Although this is not pursued further in this work, we would like to stress that this finding shows that defining a contact as `Ohmic', in the sense that it does not limit injection and extraction in a particular experiment, is insufficient. Here, injection barriers of~0.1 and~\unit[0.2]{eV} both give rise to `Ohmic' injection, implying bulk-limited transport under forward bias, but these barriers are not equivalent in terms of the resulting photovoltaic behavior.

Another interesting observation is that, as one would expect, lowering the injection barriers from 0.2 to \unit[0.1]{eV} leads to an increase in~$\voc$. But at the same time the fill factor becomes reduced, so that the overall power conversion efficiency stays roughly the same. Hence, we can deduce from our KMC~simulations that reducing the injection barriers does not \textit{per se} lead to a better performing solar cell device. Closer examination of this aspect, however, requires more extensive parameter studies, which are beyond the scope of the present paper and will be the subject of future work.

\subsection{Morphology Governs Charge Recombination}

\begin{figure*}[t]
\includegraphics{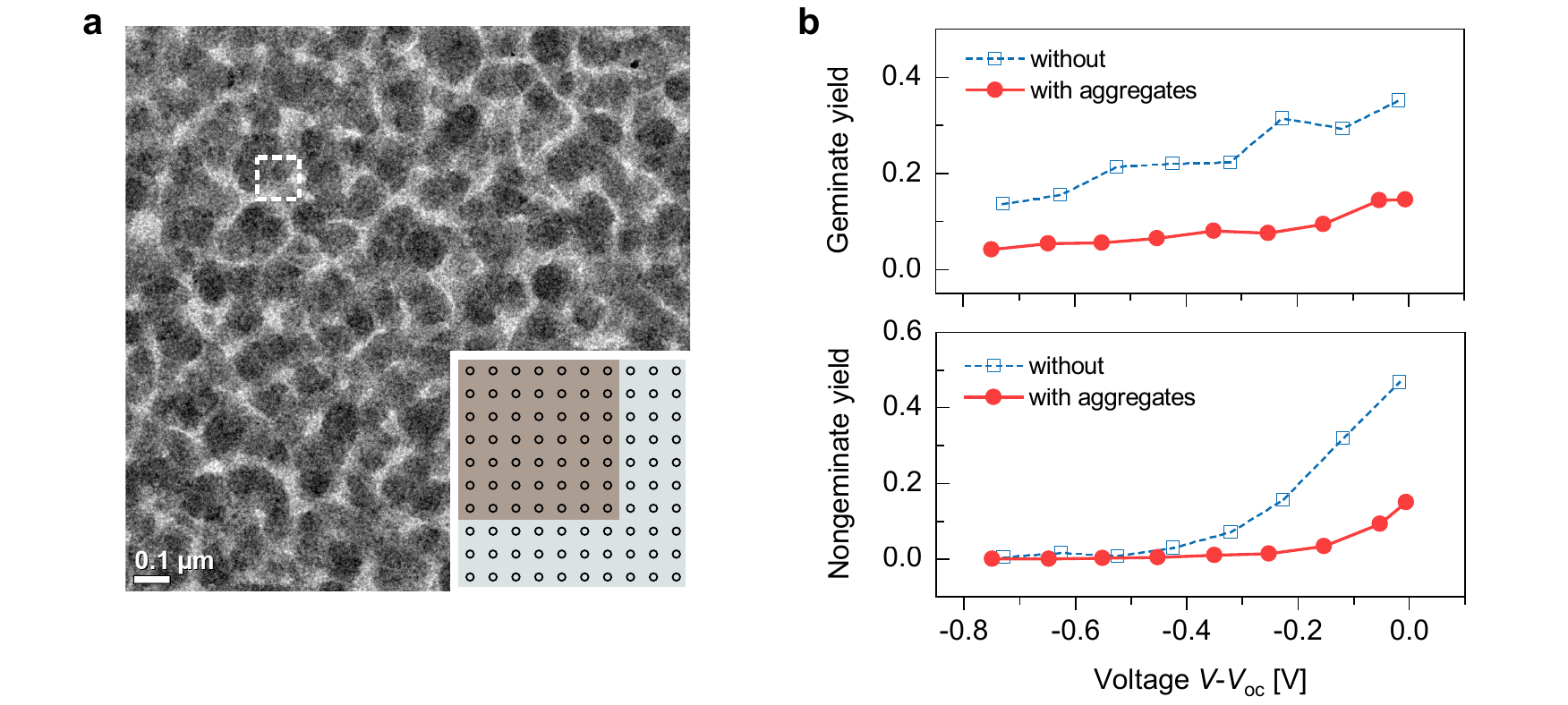}
\caption{Impact of morphology on charge recombination. (a)~BF-TEM~image of a TQ1:P\ce{C71}BM blend film and numerical implementation of the morphology~(inset). See the Supporting Information for a HDAAF-STEM image of the same sample. The relevant structure are P\ce{C71}BM aggregates, which are assumed in the KMC model as $7 \times 7$~inclusions in a $10 \times 10$~unit cell representing the mixed phase of TQ1 and P\ce{C71}BM. In vertical direction, the aggregates are assumed to be extended throughout the whole thickness. The dashed square shows a region that is reasonably captured by this model morphology. (b)~Impact of the aggregates on the simulated yield of geminate and nongeminate recombination between short-circuit and open-circuit conditions.}
\label{fig:figure3}
\end{figure*}

In our previous KMC~studies the photoactive blend was assumed as an effective hopping medium without any morphological features.\cite{Melianas2015,Melianas2014,Melianas2017} This zero-order approximation is reasonable when describing experiments on the ps--$\mu$s timescale where charge recombination is insignificant. However, we find that the effective-medium approach fails to fully describe the device $J$--$V$ curve (Figure~\ref{fig:figure1}, open squares). In order to obtain a more realistic picture of the morphology, we performed transmission electron microscopy~(TEM). Figure~\ref{fig:figure3}a shows a representative bright-field~(BF) TEM~image of a TQ1:P\ce{C71}BM blend that was prepared the same way as for device fabrication. The image displays a granular structure with clusters of dark contrast of about~\unit[100]{nm} in size. Dark regions in BF-TEM images of polymer/fullerene blends are commonly attributed to fullerene domains because of their higher density. However, this assignment is not unambiguous; the different intensities could also be caused by phase contrast due to local crystallinity differences. For comparison, we investigated the same sample in scanning transmission electron microscopy~(STEM) mode using a high-angle annular dark field~(HAADF) detector.\cite{Loos2009,Alekseev2015} In the Supporting Information we show that HAADF-STEM reveals very similar structures as in Figure~\ref{fig:figure3}a, but of inverted contrast. This clearly confirms that the clusters seen in TEM are P\ce{C71}BM~aggregates, in agreement with earlier work on similar blend systems.\cite{Backe2015}

The main effect of aggregation is to reduce the energy gap between the highest occupied molecular orbital~(HOMO) and the lowest unoccupied molecular orbital~(LUMO) compared to the amorphous material. This creates an energy cascade with a driving force for carriers to move from the (molecularly mixed)~amorphous regions towards the (material-pure)~aggregates and will affect the way how charges separate and recombine.\cite{Burke2014,McMahon2011,Sweetnam2014,Jamieson2012} We implemented the aggregates in the KMC~model as $7 \times 7$~inclusions in a $10 \times 10$~unit cell describing the mixed donor/acceptor phase~(Figure~\ref{fig:figure3}a, inset). Inclusions were assumed to consist of pure P\ce{C71}BM with a \unit[0.2]{eV} lower-lying LUMO compared to the mixed phase; all other properties were left unchanged to keep the number of unknown parameters at a minimum. We did not consider pure TQ1 domains, as our TEM experiments do not provide any evidence for them. This is reasonable, since TQ1 is a relatively amorphous polymer that has no strong tendency to form aggregates, in particular in blends with excess fullerene.\cite{Wang2013}

Note that the aggregate size in the KMC~model is smaller than what is suggested from the electron microscopy images. This was done to keep the simulation box computationally tractable while still getting reasonable statistics. The size of the inclusions and the unit cell were chosen such that the donor/acceptor ratio of the blend is maintained. A detailed examination of the structure size on the device performance is beyond the scope of this work; however, first tests indicate that the actual size of the aggregates is much less important than their presence. Likewise, a \unit[0.1]{eV} lower-lying LUMO for the aggregate phase did not make any relevant difference as compared to the used~\unit[0.2]{eV}.

Only with the inclusions in the effective hopping medium we were able to match the fill factor of the experimental devices. Figure~\ref{fig:figure3}b shows that this is due to a reduction of the charge recombination. Importantly, the presence of aggregates simultaneously reduces the yields of geminate and nongeminate recombination. This confirms earlier suggestions that the generation and recombination of free charges are coupled via the ability of CT pairs to separate.\cite{Burke2015,Shoaee2019} In other words, the possibility for carriers~(here: electrons) to lower their energy by moving to the aggregates will not only increase the charge separation yield, but also reduce the nongeminate recombination. This is a clear hint that the different ability to form aggregates/phase-pure domains may explain why different OPV~materials show so different recombination rates compared to the Langevin model. In the context of this work, however, it means that it is the kinetics of the CT~states, i.e., how they dissociate and (re-)associate, that must be calibrated experimentally.

\subsection{Calibration of the Recombination Rate}

\begin{figure*}[t]
\includegraphics{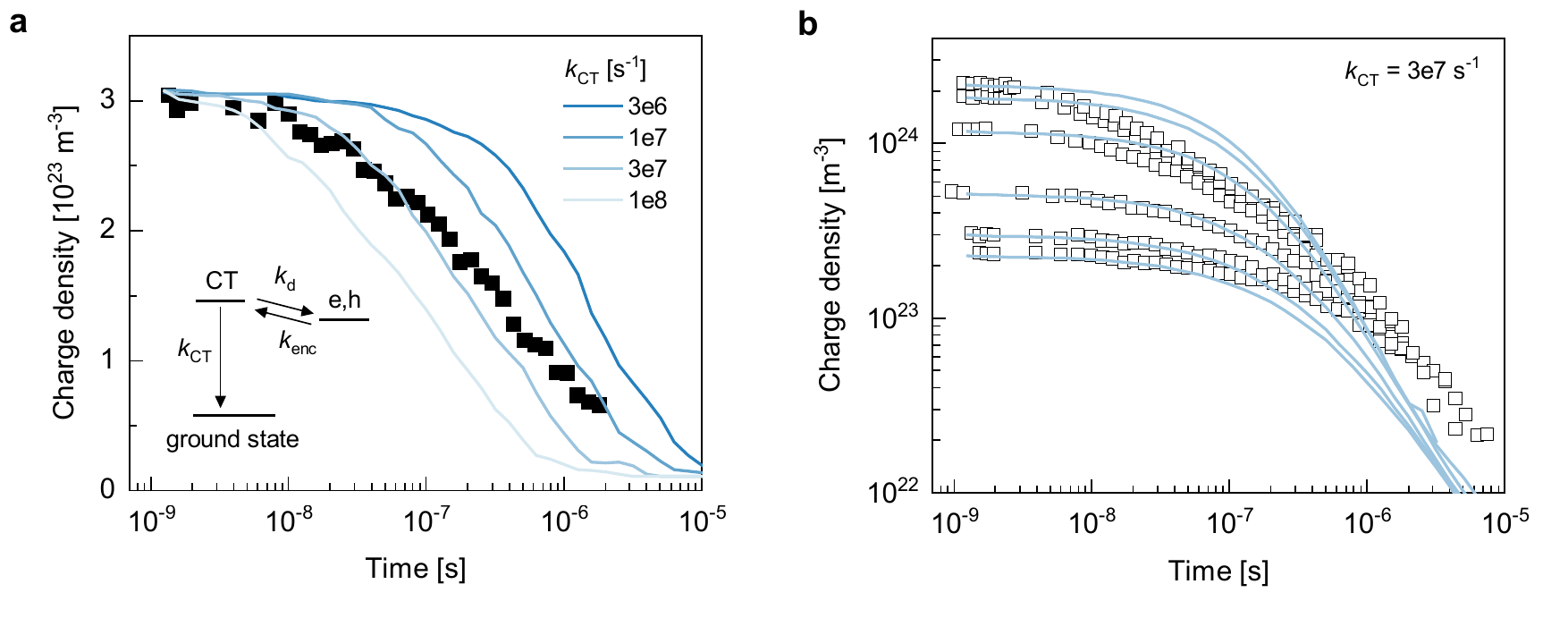}
\caption{Comparison between experimental~(symbols) and modeled~(lines) transient absorption of TQ1:P\ce{C71}BM blends. (a)~Attempts to model the experiment with a pump fluency of~$\unit[4 \times 10^{16}]{photons \cdot m^{-2}}$ with the parameters in Table~\ref{tab:params} but varied decay rate of CT~states into the ground state. Inset:~Illustration of the relevant states and transitions for charge recombination. (b)~Measurements with various initial carrier densities by varying the pump fluency from $2.5 \times 10^{16}$ to $\unit[6 \times 10^{17}]{photons \cdot m^{-2}}$ and simulations for a fixed decay rate of~$\kct = \unit[3 \times 10^7]{s^{-1}}$. Experimental data from Andersson~et~al.\cite{Andersson2013}}
\label{fig:figure4}
\end{figure*}

The inset in Figure~\ref{fig:figure4} illustrates the kinetic model of charge recombination that has emerged from literature.\cite{Burke2015,Burke2014,Murthy2013,Shoaee2019,Howard2014} As we discuss to some detail in the Supporting Information, recombination in TQ1:P\ce{C71}BM is not limited by the rate~$\kenc$ at which free carriers meet to form an interfacial CT~complex. This implies that the probability for the CT~pair to dissociate is much higher than to decay to the ground state~($\kd \gg \kct$). It has been shown that in such a situation an equilibrium between CT~states and free charge carriers is established.\cite{Burke2015,Howard2014} The position of the equilibrium is determined by the rate~$\kct$, which is the relevant parameter in the KMC~model to calibrate the recombination.

In order to do so, we use the results of transient absorption~(TA) experiments. TA~is a pump-probe technique that optically tracks a carrier population created by a short light pulse over time. As the experiment is carried out under flat-band conditions, the measured decay solely reflects the recombination kinetics. Figure~\ref{fig:figure4}a shows the TA~decay of a TQ1:P\ce{C71}BM~device for a pump fluency of $\unit[4 \times 10^{16}]{photons \cdot m^{-2}}$ taken from literature.\cite{Andersson2013} The traces are attempts to describe the experiment with our KMC model. One can clearly see that the (inverse) CT~state lifetime is the crucial parameter for the decay dynamics. The best fit on short time scales is obtained for $\kct = \unit[3 \times 10^7]{s^{-1}}$. Figure~\ref{fig:figure4}b demonstrates that with the calibrated value for~$\kct$, we are able to reasonably describe transient absorption data for a range of initial carrier densities.

On longer time scales, however, the fit between TA~experiment and KMC~model is not as good. The reason for this are the symmetrized transport parameters we use for computational effectiveness. As discussed in the Supporting Information, the disorder~$\sigma$ and attempt-to-hop frequency~$\nu_0$ are largely interchangeable, i.e., increasing the one parameter can be compensated by decreasing the other and vice versa. This interchangeability allows us to use the values given in Table~\ref{tab:params}, which keep the KMC~calculations manageable, while still reproducing the measured quasi-steady state mobilities. Nevertheless, using symmetrized transport parameters remains a simplification, so that some of the details necessary to describe the full TA~traces are lost. In Figure~S2 we show that a better fit can be obtained when the `real', non-symmetrized values for~$\sigma$ and $\nu_0$ are used in the simulation. However, significant differences between the parameter sets are only noticeable at very high initial carrier densities~($\sim\unit[10^{24}]{m^{-3}}$) and on the time scale of~$\mu$s and beyond. At those times, most of the carriers have already been extracted, as can be seen from the histograms in Figure~\ref{fig:figure3}b and from previous experiments.\cite{Melianas2014,Melianas2017,Melianas2019} Hence, the use of the simplified transport parameters is well justified when describing a solar cell under standard operating conditions.

\section{Conclusions}
We have presented a KMC model that fully describes the $J$--$V$ curve of a disordered organic solar cell under solar illumination. The agreement between experiment and simulation is obtained by experimentally calibrating the injections barriers, the blend morphology, and the dynamics of the CT~state involved in charge recombination. Our work clearly highlights the importance of contacts for a KMC~model to describe operating OPVs. We find that seemingly small changes in the injection barrier height can have major impact on the device~$\voc$ and fill factor. This confirms that injected charges play a key role in the apparent recombination mechanism. Furthermore, we find charge recombination to be limited by the fate of the intermediate CT~exciton, which can be influenced by the presence of aggregates in the active layer, and not by the transport of electrons and holes; our results indicate that the energy difference between the aggregated and mixed regions and the aggregate size is not that important, but the presence of aggregates is.

The platform introduced in this work will be useful for future studies on properties of OPV~materials that are not accessible via macroscopic, quasi-equilibrium modeling techniques such as drift--diffusion. Questions to be answered include but are not limited to how nonequilibrium effects affect the device operation and what the critical morphological factors are that determine the charge recombination. Finally, we point out that our results give strong support that the conclusions derived from previous transient KMC studies are also relevant for OPVs under standard operating conditions.

\section{Experimental Section}

\paragraph{Device Fabrication:} Binary solution of poly[[2,3-bis(3-octyl\-oxy\-phenyl)-5,8-quin\-oxaline\-diyl]-2,5-thio\-phene\-diyl]~(TQ1) and [6,6]-phenyl-\ce{C71}-butyric acid methyl ester~(P\ce{C71}BM) in  weight ratio 1:2.5 was prepared in chlorobenzene to a total concentration of~$\unit[25]{mg\,mL^{-1}}$. The device structure was ITO\slash{}PEDOT:PSS (\unit[30]{nm})\slash{}TQ1:P\ce{C71}BM (\unit[72]{nm})\slash{}LiF (\unit[0.6]{nm})\slash{}Al (\unit[90]{nm}). ITO-coated glass substrates were boiled in a 5:1:1 mixture~(by volume) of deionized water, ammonium hydroxide~(25\%) and hydrogen peroxide~(28\%) at $\unit[80]{^\circ C}$ for \unit[15]{min} for cleaning. PEDOT:PSS~(Baytron P VP Al 4083) was spin-coated onto the ITO glasses at~\unit[3000]{rpm} for~\unit[40]{s}, followed by annealing at~$\unit[150]{^\circ C}$ for~\unit[10]{min}. The active layer was spin coated at~\unit[500]{rpm} for~\unit[60]{s}. The LiF/Al top electrode was deposited by thermal evaporation through a shadow mask to get an active area of~$\unit[0.05]{cm^2}$. 

\paragraph{Electrical Measurements:} Current--voltage curves were recorded with a Keithley 2401 source measure unit under standard AM1.5G illumination~($\unit[100]{mWcm^{-2}}$) using an Oriel LSH-7320 solar simulator. Dark charge extraction measurements were performed using a pulse generator~(SRS DG~535) and a function generator~(SRS DS~345) for applying the extraction voltage pulse and an oscilloscope~(Tektronix TDS 680B) for recording the current transient. Devices were mounted in a vacuum cryostat kept at room temperature. The measurement setup 
was controlled from a computer using a LabVIEW program. In the CELIV experiments, a steady-state voltage~$\vpre$ was applied in forward bias of the solar cell and a linearly increasing extraction pulse~$V(t) = -At$ with~$A = \unit[0.05]{V\,\mu s^{-1}}$ and a total pulse length of~$\unit[50]{\mu s}$ was used for charge extraction. For the BACE measurements, the same~$\vpre$ as used in the CELIV measurements was applied and charges were extracted using a rectangular voltage pulse with an amplitude~\unit[2.5]{V} and a pulse length of~$\unit[50]{\mu s}$.

\paragraph{Electron Microscopy:} Samples for TEM were prepared by floating off TQ1:P\ce{C71}BM~films from PEDOT:PSS-coated glass substrates in deionized water. This was followed by picking up the films directly on TEM~copper mesh grids for imaging. BF-TEM~images were taken at an acceleration voltage of~\unit[200]{kV} in a FEI~Tecnai~T20 instrument. HAADF-STEM images were taken at an acceleration voltage of~\unit[300]{kV} in a FEI Titan~80-300.

\begin{acknowledgement}
This project has received funding through European Union's Horizon 2020 research and innovation programme under the Marie Sk\l{}odowska-Curie grant agreement No~799801 (`ReMorphOPV'). T.U., G.P. and E.O. acknowledges funding by Vetenskapsr\aa{}det~(project `OPV2.0'). A.M. acknowledges support from the Knut and Alice Wallenberg Foundation~(KAW 2016.0494) for Postdoctoral Research at Stanford University.  S.D. and R.\"O. acknowledge financial support from the Jane \& Aatos Erkko foundation (project `ASPIRE'). G.P. and E.O. thank the Chalmers Material Analysis Laboratory for their support of microscopes.
\end{acknowledgement}

\section*{Keywords}
Organic Photovoltaics, Kinetic Monte Carlo Simulations, Charge Injection, Charge Recombination, Morphology

\bibliography{ms}

\providecommand{\latin}[1]{#1}
\makeatletter
\providecommand{\doi}
  {\begingroup\let\do\@makeother\dospecials
  \catcode`\{=1 \catcode`\}=2\doi@aux}
\providecommand{\doi@aux}[1]{\endgroup\texttt{#1}}
\makeatother
\providecommand*\mcitethebibliography{\thebibliography}
\csname @ifundefined\endcsname{endmcitethebibliography}
  {\let\endmcitethebibliography\endthebibliography}{}
\begin{mcitethebibliography}{44}
\providecommand*\natexlab[1]{#1}
\providecommand*\mciteSetBstSublistMode[1]{}
\providecommand*\mciteSetBstMaxWidthForm[2]{}
\providecommand*\mciteBstWouldAddEndPuncttrue
  {\def\EndOfBibitem{\unskip.}}
\providecommand*\mciteBstWouldAddEndPunctfalse
  {\let\EndOfBibitem\relax}
\providecommand*\mciteSetBstMidEndSepPunct[3]{}
\providecommand*\mciteSetBstSublistLabelBeginEnd[3]{}
\providecommand*\EndOfBibitem{}
\mciteSetBstSublistMode{f}
\mciteSetBstMaxWidthForm{subitem}{(\alph{mcitesubitemcount})}
\mciteSetBstSublistLabelBeginEnd
  {\mcitemaxwidthsubitemform\space}
  {\relax}
  {\relax}

\bibitem[Melianas \latin{et~al.}(2014)Melianas, Pranculis, Devi\v{z}is,
  Gulbinas, Ingan\"{a}s, and Kemerink]{Melianas2014}
Melianas,~A.; Pranculis,~V.; Devi\v{z}is,~A.; Gulbinas,~V.; Ingan\"{a}s,~O.;
  Kemerink,~M. {Dispersion-Dominated Photocurrent in Polymer:Fullerene Solar
  Cells}. \emph{Adv. Funct. Mater.} \textbf{2014}, \emph{24}, 4507--4514\relax
\mciteBstWouldAddEndPuncttrue
\mciteSetBstMidEndSepPunct{\mcitedefaultmidpunct}
{\mcitedefaultendpunct}{\mcitedefaultseppunct}\relax
\EndOfBibitem
\bibitem[Howard \latin{et~al.}(2014)Howard, Etzold, Laquai, and
  Kemerink]{Howard2014}
Howard,~I.~A.; Etzold,~F.; Laquai,~F.; Kemerink,~M. {Nonequilibrium Charge
  Dynamics in Organic Solar Cells}. \emph{Adv. Energy Mater.} \textbf{2014},
  \emph{4}, 1301743\relax
\mciteBstWouldAddEndPuncttrue
\mciteSetBstMidEndSepPunct{\mcitedefaultmidpunct}
{\mcitedefaultendpunct}{\mcitedefaultseppunct}\relax
\EndOfBibitem
\bibitem[Melianas \latin{et~al.}(2015)Melianas, Etzold, Savenije, Laquai,
  Ingan\"{a}s, and Kemerink]{Melianas2015}
Melianas,~A.; Etzold,~F.; Savenije,~T.~J.; Laquai,~F.; Ingan\"{a}s,~O.;
  Kemerink,~M. {Photo-Generated Carriers Lose Energy During Extraction from
  Polymer-Fullerene Solar Cells}. \emph{Nat. Commun.} \textbf{2015}, \emph{6},
  8778\relax
\mciteBstWouldAddEndPuncttrue
\mciteSetBstMidEndSepPunct{\mcitedefaultmidpunct}
{\mcitedefaultendpunct}{\mcitedefaultseppunct}\relax
\EndOfBibitem
\bibitem[Melianas \latin{et~al.}(2017)Melianas, Pranculis, Xia, Felekidis,
  Ingan\"{a}s, Gulbinas, and Kemerink]{Melianas2017}
Melianas,~A.; Pranculis,~V.; Xia,~Y.; Felekidis,~N.; Ingan\"{a}s,~O.;
  Gulbinas,~V.; Kemerink,~M. {Photogenerated Carrier Mobility Significantly
  Exceeds Injected Carrier Mobility in Organic Solar Cells}. \emph{Adv. Energy
  Mater.} \textbf{2017}, \emph{7}, 1602143\relax
\mciteBstWouldAddEndPuncttrue
\mciteSetBstMidEndSepPunct{\mcitedefaultmidpunct}
{\mcitedefaultendpunct}{\mcitedefaultseppunct}\relax
\EndOfBibitem
\bibitem[Lyons \latin{et~al.}(2012)Lyons, Clarke, and Groves]{Lyons2012}
Lyons,~B.~P.; Clarke,~N.; Groves,~C. The Relative Importance of Domain Size,
  Domain Purity and Domain Interfaces to the Performance of Bulk-Heterojunction
  Organic Photovoltaics. \emph{Energy Environ. Sci.} \textbf{2012}, \emph{5},
  7657--7663\relax
\mciteBstWouldAddEndPuncttrue
\mciteSetBstMidEndSepPunct{\mcitedefaultmidpunct}
{\mcitedefaultendpunct}{\mcitedefaultseppunct}\relax
\EndOfBibitem
\bibitem[Heiber \latin{et~al.}(2015)Heiber, Baumbach, Dyakonov, and
  Deibel]{Heiber2015}
Heiber,~M.~C.; Baumbach,~C.; Dyakonov,~V.; Deibel,~C. Encounter-Limited
  Charge-Carrier Recombination in Phase-Separated Organic Semiconductor Blends.
  \emph{Phys. Rev. Lett.} \textbf{2015}, \emph{114}, 136602\relax
\mciteBstWouldAddEndPuncttrue
\mciteSetBstMidEndSepPunct{\mcitedefaultmidpunct}
{\mcitedefaultendpunct}{\mcitedefaultseppunct}\relax
\EndOfBibitem
\bibitem[Groves and Greenham(2008)Groves, and Greenham]{Groves2008}
Groves,~C.; Greenham,~N.~C. {Bimolecular Recombination in Polymer Electronic
  Devices}. \emph{Phys. Rev. B} \textbf{2008}, \emph{78}, 155205\relax
\mciteBstWouldAddEndPuncttrue
\mciteSetBstMidEndSepPunct{\mcitedefaultmidpunct}
{\mcitedefaultendpunct}{\mcitedefaultseppunct}\relax
\EndOfBibitem
\bibitem[Coropceanu \latin{et~al.}(2017)Coropceanu, Br\'{e}das, and
  Mehraeen]{Coropceanu2017}
Coropceanu,~V.; Br\'{e}das,~J.~L.; Mehraeen,~S. {Impact of Active Layer
  Morphology on Bimolecular Recombination Dynamics in Organic Solar Cells}.
  \emph{J. Phys. Chem. C} \textbf{2017}, \emph{121}, 24954--24961\relax
\mciteBstWouldAddEndPuncttrue
\mciteSetBstMidEndSepPunct{\mcitedefaultmidpunct}
{\mcitedefaultendpunct}{\mcitedefaultseppunct}\relax
\EndOfBibitem
\bibitem[Kaiser and Gagliardi(2019)Kaiser, and Gagliardi]{Kaiser2019}
Kaiser,~W.; Gagliardi,~A. {Kinetic Monte Carlo Study of the Role of the
  Energetic Disorder on the Open-Circuit Voltage in Polymer/Fullerene Solar
  Cells}. \emph{J. Phys. Chem. Lett.} \textbf{2019}, \emph{10},
  6097--6104\relax
\mciteBstWouldAddEndPuncttrue
\mciteSetBstMidEndSepPunct{\mcitedefaultmidpunct}
{\mcitedefaultendpunct}{\mcitedefaultseppunct}\relax
\EndOfBibitem
\bibitem[Meng \latin{et~al.}(2010)Meng, Shang, Li, Li, Zhan, Shuai, Kimber, and
  Walker]{Meng2010}
Meng,~L.; Shang,~Y.; Li,~Q.; Li,~Y.; Zhan,~X.; Shuai,~Z.; Kimber,~R. G.~E.;
  Walker,~A.~B. {Dynamic Monte Carlo Simulation for Highly Efficient Polymer
  Blend Photovoltaics}. \emph{J. Phys. Chem. B} \textbf{2010}, \emph{114},
  36--41\relax
\mciteBstWouldAddEndPuncttrue
\mciteSetBstMidEndSepPunct{\mcitedefaultmidpunct}
{\mcitedefaultendpunct}{\mcitedefaultseppunct}\relax
\EndOfBibitem
\bibitem[Meng \latin{et~al.}(2011)Meng, Wang, Li, Yi, Br\'{e}das, and
  Shuai]{Meng2011}
Meng,~L.; Wang,~D.; Li,~Q.; Yi,~Y.; Br\'{e}das,~J.~L.; Shuai,~Z. {An Improved
  Dynamic Monte Carlo Model Coupled with Poisson Equation to Simulate the
  Performance of Organic Photovoltaic Devices}. \emph{J. Chem. Phys.}
  \textbf{2011}, \emph{134}, 124102\relax
\mciteBstWouldAddEndPuncttrue
\mciteSetBstMidEndSepPunct{\mcitedefaultmidpunct}
{\mcitedefaultendpunct}{\mcitedefaultseppunct}\relax
\EndOfBibitem
\bibitem[Kipp and Ganesan(2013)Kipp, and Ganesan]{Kipp2013}
Kipp,~D.; Ganesan,~V. {A Kinetic Monte Carlo Model with Improved Charge
  Injection Model for the Photocurrent Characteristics of Organic Solar Cells}.
  \emph{J. Appl. Phys.} \textbf{2013}, \emph{113}, 234502\relax
\mciteBstWouldAddEndPuncttrue
\mciteSetBstMidEndSepPunct{\mcitedefaultmidpunct}
{\mcitedefaultendpunct}{\mcitedefaultseppunct}\relax
\EndOfBibitem
\bibitem[W\"{u}rfel \latin{et~al.}(2019)W\"{u}rfel, Perdig\'{o}n-Toro,
  Kurpiers, Wolff, Caprioglio, Rech, Zhu, Zhan, You, Shoaee, Neher, and
  Stolterfoht]{Wurfel2019}
W\"{u}rfel,~U.; Perdig\'{o}n-Toro,~L.; Kurpiers,~J.; Wolff,~C.~M.;
  Caprioglio,~P.; Rech,~J.~J.; Zhu,~J.; Zhan,~X.; You,~W.; Shoaee,~S.;
  Neher,~D.; Stolterfoht,~M. Recombination between Photogenerated and
  Electrode-Induced Charges Dominates the Fill Factor Losses in Optimized
  Organic Solar Cells. \emph{J. Phys. Chem. Lett.} \textbf{2019}, \emph{10},
  3473--3480\relax
\mciteBstWouldAddEndPuncttrue
\mciteSetBstMidEndSepPunct{\mcitedefaultmidpunct}
{\mcitedefaultendpunct}{\mcitedefaultseppunct}\relax
\EndOfBibitem
\bibitem[Neher \latin{et~al.}(2016)Neher, Kniepert, Elimelech, and
  Koster]{Neher2016}
Neher,~D.; Kniepert,~J.; Elimelech,~A.; Koster,~L. J.~A. {A New Figure of Merit
  for Organic Solar Cells with Transport-limited Photocurrents}. \emph{Sci.
  Rep.} \textbf{2016}, \emph{6}, 24861\relax
\mciteBstWouldAddEndPuncttrue
\mciteSetBstMidEndSepPunct{\mcitedefaultmidpunct}
{\mcitedefaultendpunct}{\mcitedefaultseppunct}\relax
\EndOfBibitem
\bibitem[Bartesaghi \latin{et~al.}(2015)Bartesaghi, {Del Carmen P\'{e}rez},
  Kniepert, Roland, Turbiez, Neher, and Koster]{Bartesaghi2015}
Bartesaghi,~D.; {Del Carmen P\'{e}rez},~I.; Kniepert,~J.; Roland,~S.;
  Turbiez,~M.; Neher,~D.; Koster,~L. J.~A. {Competition Between Recombination
  and Extraction of Free Charges Determines the Fill Factor of Organic Solar
  Cells}. \emph{Nat. Commun.} \textbf{2015}, \emph{6}, 7083\relax
\mciteBstWouldAddEndPuncttrue
\mciteSetBstMidEndSepPunct{\mcitedefaultmidpunct}
{\mcitedefaultendpunct}{\mcitedefaultseppunct}\relax
\EndOfBibitem
\bibitem[Kaienburg \latin{et~al.}(2016)Kaienburg, Rau, and
  Kirchartz]{Kaienburg2016}
Kaienburg,~P.; Rau,~U.; Kirchartz,~T. {Extracting Information about the
  Electronic Quality of Organic Solar-Cell Absorbers from Fill Factor and
  Thickness}. \emph{Phys. Rev. Appl.} \textbf{2016}, \emph{6}, 024001\relax
\mciteBstWouldAddEndPuncttrue
\mciteSetBstMidEndSepPunct{\mcitedefaultmidpunct}
{\mcitedefaultendpunct}{\mcitedefaultseppunct}\relax
\EndOfBibitem
\bibitem[G\"{o}hler \latin{et~al.}(2018)G\"{o}hler, Wagenpfahl, and
  Deibel]{Gohler2018}
G\"{o}hler,~C.; Wagenpfahl,~A.; Deibel,~C. {Nongeminate Recombination in
  Organic Solar Cells}. \emph{Adv. Electron. Mater.} \textbf{2018}, \emph{4},
  1700505\relax
\mciteBstWouldAddEndPuncttrue
\mciteSetBstMidEndSepPunct{\mcitedefaultmidpunct}
{\mcitedefaultendpunct}{\mcitedefaultseppunct}\relax
\EndOfBibitem
\bibitem[Lakhwani \latin{et~al.}(2014)Lakhwani, Rao, and Friend]{Lakhwani2014}
Lakhwani,~G.; Rao,~A.; Friend,~R.~H. Bimolecular Recombination in Organic
  Photovoltaics. \emph{Annu. Rev. Phys. Chem.} \textbf{2014}, \emph{65},
  557--581\relax
\mciteBstWouldAddEndPuncttrue
\mciteSetBstMidEndSepPunct{\mcitedefaultmidpunct}
{\mcitedefaultendpunct}{\mcitedefaultseppunct}\relax
\EndOfBibitem
\bibitem[Burke and McGehee(2014)Burke, and McGehee]{Burke2014}
Burke,~T.~M.; McGehee,~M.~D. {How High Local Charge Carrier Mobility and an
  Energy Cascade in a Three-Phase Bulk Heterojunction Enable \textgreater90\%
  Quantum Efficiency}. \emph{Adv. Mater.} \textbf{2014}, \emph{26},
  1923--1928\relax
\mciteBstWouldAddEndPuncttrue
\mciteSetBstMidEndSepPunct{\mcitedefaultmidpunct}
{\mcitedefaultendpunct}{\mcitedefaultseppunct}\relax
\EndOfBibitem
\bibitem[Jamieson \latin{et~al.}(2012)Jamieson, Domingo, McCarthy-Ward, Heeney,
  Stingelin, and Durrant]{Jamieson2012}
Jamieson,~F.~C.; Domingo,~E.~B.; McCarthy-Ward,~T.; Heeney,~M.; Stingelin,~N.;
  Durrant,~J.~R. {Fullerene Crystallisation as a Key Driver of Charge
  Separation in Polymer/Fullerene Bulk Heterojunction Solar Cells}. \emph{Chem.
  Sci.} \textbf{2012}, \emph{3}, 485--492\relax
\mciteBstWouldAddEndPuncttrue
\mciteSetBstMidEndSepPunct{\mcitedefaultmidpunct}
{\mcitedefaultendpunct}{\mcitedefaultseppunct}\relax
\EndOfBibitem
\bibitem[Sweetnam \latin{et~al.}(2014)Sweetnam, Graham, Ngongang~Ndjawa,
  Heum\"uller, Bartelt, Burke, Li, You, Amassian, and McGehee]{Sweetnam2014}
Sweetnam,~S.; Graham,~K.~R.; Ngongang~Ndjawa,~G.~O.; Heum\"uller,~T.;
  Bartelt,~J.~A.; Burke,~T.~M.; Li,~W.; You,~W.; Amassian,~A.; McGehee,~M.~D.
  {Characterization of the Polymer Energy Landscape in Polymer:Fullerene Bulk
  Heterojunctions with Pure and Mixed Phases}. \emph{J. Am. Chem. Soc.}
  \textbf{2014}, \emph{136}, 14078--14088\relax
\mciteBstWouldAddEndPuncttrue
\mciteSetBstMidEndSepPunct{\mcitedefaultmidpunct}
{\mcitedefaultendpunct}{\mcitedefaultseppunct}\relax
\EndOfBibitem
\bibitem[McMahon \latin{et~al.}(2011)McMahon, Cheung, and Troisi]{McMahon2011}
McMahon,~D.~P.; Cheung,~D.~L.; Troisi,~A. {Why Holes and Electrons Separate So
  Well in Polymer/Fullerene Photovoltaic Cells}. \emph{J. Phys. Chem. Lett.}
  \textbf{2011}, \emph{2}, 2737--2741\relax
\mciteBstWouldAddEndPuncttrue
\mciteSetBstMidEndSepPunct{\mcitedefaultmidpunct}
{\mcitedefaultendpunct}{\mcitedefaultseppunct}\relax
\EndOfBibitem
\bibitem[Melianas and Kemerink(2019)Melianas, and Kemerink]{Melianas2019}
Melianas,~A.; Kemerink,~M. {Photogenerated Charge Transport in Organic
  Electronic Materials: Experiments Confirmed by Simulations}. \emph{Adv.
  Mater.} \textbf{2019}, 1806004\relax
\mciteBstWouldAddEndPuncttrue
\mciteSetBstMidEndSepPunct{\mcitedefaultmidpunct}
{\mcitedefaultendpunct}{\mcitedefaultseppunct}\relax
\EndOfBibitem
\bibitem[Wang \latin{et~al.}(2010)Wang, Hou, Wang, Hellstr\"{o}m, Zhang,
  Ingan\"{a}s, and Andersson]{Wang2010}
Wang,~E.; Hou,~L.; Wang,~Z.; Hellstr\"{o}m,~S.; Zhang,~F.; Ingan\"{a}s,~O.;
  Andersson,~M.~R. {An Easily Synthesized Blue Polymer for High‐Performance
  Polymer Solar Cells}. \emph{Adv. Mater.} \textbf{2010}, \emph{22},
  5240--5244\relax
\mciteBstWouldAddEndPuncttrue
\mciteSetBstMidEndSepPunct{\mcitedefaultmidpunct}
{\mcitedefaultendpunct}{\mcitedefaultseppunct}\relax
\EndOfBibitem
\bibitem[Melianas \latin{et~al.}(2019)Melianas, Felekidis, Puttisong, Meskers,
  Ingan\"{a}s, Chen, and Kemerink]{Melianas2019b}
Melianas,~A.; Felekidis,~N.; Puttisong,~Y.; Meskers,~S. C.~J.; Ingan\"{a}s,~O.;
  Chen,~W.~M.; Kemerink,~M. {Nonequilibrium Site Distribution Governs
  Charge-Transfer Electroluminescence at Disordered Organic Heterointerfaces}.
  \emph{Proc. Natl. Acad. Sci. USA} \textbf{2019}, \emph{116},
  23416--23425\relax
\mciteBstWouldAddEndPuncttrue
\mciteSetBstMidEndSepPunct{\mcitedefaultmidpunct}
{\mcitedefaultendpunct}{\mcitedefaultseppunct}\relax
\EndOfBibitem
\bibitem[Upreti \latin{et~al.}(2019)Upreti, Wang, Zhang, Scheunemann, Gao, and
  Kemerink]{Upreti2019}
Upreti,~T.; Wang,~Y.; Zhang,~H.; Scheunemann,~D.; Gao,~F.; Kemerink,~M.
  {Experimentally Validated Hopping-Transport Model for Energetically
  Disordered Organic Semiconductors}. \emph{Phys. Rev. Applied} \textbf{2019},
  \emph{12}, 064039\relax
\mciteBstWouldAddEndPuncttrue
\mciteSetBstMidEndSepPunct{\mcitedefaultmidpunct}
{\mcitedefaultendpunct}{\mcitedefaultseppunct}\relax
\EndOfBibitem
\bibitem[Burgelman \latin{et~al.}(2000)Burgelman, Nollet, and
  Degrave]{Burgelman2000}
Burgelman,~M.; Nollet,~P.; Degrave,~S. {Modelling Polycrystalline Semiconductor
  Solar Cells}. \emph{Thin Solid Films} \textbf{2000}, \emph{361-362},
  527--532\relax
\mciteBstWouldAddEndPuncttrue
\mciteSetBstMidEndSepPunct{\mcitedefaultmidpunct}
{\mcitedefaultendpunct}{\mcitedefaultseppunct}\relax
\EndOfBibitem
\bibitem[Wilken \latin{et~al.}(2020)Wilken, Sandberg, Scheunemann, and
  \"{O}sterbacka]{Wilken2020}
Wilken,~S.; Sandberg,~O.~J.; Scheunemann,~D.; \"{O}sterbacka,~R. {Watching
  Space Charge Build up in an Organic Solar Cell}. \emph{Sol. RRL}
  \textbf{2020}, \emph{4}, 1900505\relax
\mciteBstWouldAddEndPuncttrue
\mciteSetBstMidEndSepPunct{\mcitedefaultmidpunct}
{\mcitedefaultendpunct}{\mcitedefaultseppunct}\relax
\EndOfBibitem
\bibitem[Pasveer \latin{et~al.}(2005)Pasveer, Cottaar, Tanase, Coehoorn,
  Bobbert, Blom, {de Leeuw}, and Michels]{Pasveer2005}
Pasveer,~W.~F.; Cottaar,~J.; Tanase,~C.; Coehoorn,~R.; Bobbert,~P.~A.; Blom,~P.
  W.~M.; {de Leeuw},~D.~M.; Michels,~M. A.~J. Unified Description of
  Charge-Carrier Mobilities in Disordered Semiconducting Polymers. \emph{Phys.
  Rev. Lett.} \textbf{2005}, \emph{94}, 206601\relax
\mciteBstWouldAddEndPuncttrue
\mciteSetBstMidEndSepPunct{\mcitedefaultmidpunct}
{\mcitedefaultendpunct}{\mcitedefaultseppunct}\relax
\EndOfBibitem
\bibitem[Murthy \latin{et~al.}(2013)Murthy, Melianas, Tang, Ju\v{s}ka,
  Arlauskas, Zhang, Siebbeles, Ingan\"as, and Savenije]{Murthy2013}
Murthy,~D. H.~K.; Melianas,~A.; Tang,~Z.; Ju\v{s}ka,~G.; Arlauskas,~K.;
  Zhang,~F.; Siebbeles,~L. D.~A.; Ingan\"as,~O.; Savenije,~T.~J. Origin of
  Reduced Bimolecular Recombination in Blends of Conjugated Polymers and
  Fullerenes. \emph{Adv. Funct. Mater.} \textbf{2013}, \emph{23},
  4262--4268\relax
\mciteBstWouldAddEndPuncttrue
\mciteSetBstMidEndSepPunct{\mcitedefaultmidpunct}
{\mcitedefaultendpunct}{\mcitedefaultseppunct}\relax
\EndOfBibitem
\bibitem[Roland \latin{et~al.}(2019)Roland, Kniepert, Love, Negi, Liu, Bobbert,
  Melianas, Kemerink, Hofacker, and Neher]{Roland2019}
Roland,~S.; Kniepert,~J.; Love,~J.~A.; Negi,~V.; Liu,~F.; Bobbert,~P.;
  Melianas,~A.; Kemerink,~M.; Hofacker,~A.; Neher,~D. {Equilibrated Charge
  Carrier Populations Govern Steady-State Nongeminate Recombination in
  Disordered Organic Solar Cells}. \emph{J. Phys. Chem. Lett.} \textbf{2019},
  \emph{10}, 1374--1381\relax
\mciteBstWouldAddEndPuncttrue
\mciteSetBstMidEndSepPunct{\mcitedefaultmidpunct}
{\mcitedefaultendpunct}{\mcitedefaultseppunct}\relax
\EndOfBibitem
\bibitem[Kirchartz and Nelson(2012)Kirchartz, and Nelson]{Kirchartz2012}
Kirchartz,~T.; Nelson,~J. {Meaning of Reaction Orders in Polymer:Fullerene
  Solar Cells}. \emph{Phys. Rev. B} \textbf{2012}, \emph{86}, 165201\relax
\mciteBstWouldAddEndPuncttrue
\mciteSetBstMidEndSepPunct{\mcitedefaultmidpunct}
{\mcitedefaultendpunct}{\mcitedefaultseppunct}\relax
\EndOfBibitem
\bibitem[Deledalle \latin{et~al.}(2014)Deledalle, Tuladhar, Nelson, Durrant,
  and Kirchartz]{Deledalle2014}
Deledalle,~F.; Tuladhar,~P.~S.; Nelson,~J.; Durrant,~J.~R.; Kirchartz,~T.
  {Understanding the Apparent Charge Density Dependence of Mobility and
  Lifetime in Organic Bulk Heterojunction Solar Cells}. \emph{J. Phys. Chem. C}
  \textbf{2014}, \emph{118}, 8837--8842\relax
\mciteBstWouldAddEndPuncttrue
\mciteSetBstMidEndSepPunct{\mcitedefaultmidpunct}
{\mcitedefaultendpunct}{\mcitedefaultseppunct}\relax
\EndOfBibitem
\bibitem[Scheunemann \latin{et~al.}(2019)Scheunemann, Wilken, Sandberg,
  \"{O}sterbacka, and Schiek]{Scheunemann2019}
Scheunemann,~D.; Wilken,~S.; Sandberg,~O.~J.; \"{O}sterbacka,~R.; Schiek,~M.
  {Effect of Imbalanced Charge Transport on the Interplay of Surface and Bulk
  Recombination in Organic Solar Cells}. \emph{Phys. Rev. Appl.} \textbf{2019},
  \emph{11}, 054090\relax
\mciteBstWouldAddEndPuncttrue
\mciteSetBstMidEndSepPunct{\mcitedefaultmidpunct}
{\mcitedefaultendpunct}{\mcitedefaultseppunct}\relax
\EndOfBibitem
\bibitem[Hawks \latin{et~al.}(2015)Hawks, Finck, and Schwartz]{Hawks2015}
Hawks,~S.~A.; Finck,~B.~Y.; Schwartz,~B.~J. {Theory of Current Transients in
  Planar Semiconductor Devices: Insights and Applications to Organic Solar
  Cells}. \emph{Phys. Rev. Appl.} \textbf{2015}, \emph{3}, 044014\relax
\mciteBstWouldAddEndPuncttrue
\mciteSetBstMidEndSepPunct{\mcitedefaultmidpunct}
{\mcitedefaultendpunct}{\mcitedefaultseppunct}\relax
\EndOfBibitem
\bibitem[Kniepert \latin{et~al.}(2014)Kniepert, Lange, {van der Kaap}, Koster,
  and Neher]{Kniepert2014}
Kniepert,~J.; Lange,~I.; {van der Kaap},~N.~J.; Koster,~L. J.~A.; Neher,~D. {A
  Conclusive View on Charge Generation, Recombination, and Extraction in
  As-Prepared and Annealed P3HT:PCBM Blends: Combined Experimental and
  Simulation Work}. \emph{Adv. Energy Mater.} \textbf{2014}, \emph{4},
  1301401\relax
\mciteBstWouldAddEndPuncttrue
\mciteSetBstMidEndSepPunct{\mcitedefaultmidpunct}
{\mcitedefaultendpunct}{\mcitedefaultseppunct}\relax
\EndOfBibitem
\bibitem[Loos \latin{et~al.}(2009)Loos, Sourty, Lu, {de With}, and {van
  Bavel}]{Loos2009}
Loos,~J.; Sourty,~E.; Lu,~K.; {de With},~G.; {van Bavel},~S. Imaging Polymer
  Systems with High-Angle Annular Dark Field Scanning Transmission Electron
  Microscopy (HAADF-STEM). \emph{Macromol.} \textbf{2009}, \emph{42},
  2581--2586\relax
\mciteBstWouldAddEndPuncttrue
\mciteSetBstMidEndSepPunct{\mcitedefaultmidpunct}
{\mcitedefaultendpunct}{\mcitedefaultseppunct}\relax
\EndOfBibitem
\bibitem[Alekseev \latin{et~al.}(2015)Alekseev, Hedley, Al-Afeef, Ageev, and
  Samuel]{Alekseev2015}
Alekseev,~A.; Hedley,~G.~J.; Al-Afeef,~A.; Ageev,~O.~A.; Samuel,~I. D.~W.
  {Morphology and Local Electrical Properties of PTB7:PC71BM Blends}. \emph{J.
  Mater. Chem. A} \textbf{2015}, \emph{3}, 8706--8714\relax
\mciteBstWouldAddEndPuncttrue
\mciteSetBstMidEndSepPunct{\mcitedefaultmidpunct}
{\mcitedefaultendpunct}{\mcitedefaultseppunct}\relax
\EndOfBibitem
\bibitem[B\"acke \latin{et~al.}(2015)B\"acke, Lindqvist, {de Zerio Mendaza},
  Gustafsson, Wang, Andersson, M\"uller, and Olsson]{Backe2015}
B\"acke,~O.; Lindqvist,~C.; {de Zerio Mendaza},~A.~D.; Gustafsson,~S.;
  Wang,~E.; Andersson,~M.~R.; M\"uller,~C.; Olsson,~E. {Mapping Fullerene
  Crystallization in a Photovoltaic Blend: An Electron Tomography Study}.
  \emph{Nanoscale} \textbf{2015}, \emph{7}, 8451--8456\relax
\mciteBstWouldAddEndPuncttrue
\mciteSetBstMidEndSepPunct{\mcitedefaultmidpunct}
{\mcitedefaultendpunct}{\mcitedefaultseppunct}\relax
\EndOfBibitem
\bibitem[Wang \latin{et~al.}(2013)Wang, Bergqvist, Vandewal, Ma, Hou, Lundin,
  Himmelberger, Salleo, M\"{u}ller, Ingan\"{a}s, Zhang, and
  Andersson]{Wang2013}
Wang,~E.; Bergqvist,~J.; Vandewal,~K.; Ma,~Z.; Hou,~L.; Lundin,~A.;
  Himmelberger,~S.; Salleo,~A.; M\"{u}ller,~C.; Ingan\"{a}s,~O.; Zhang,~F.;
  Andersson,~M. {Conformational Disorder Enhances Solubility and Photovoltaic
  Performance of a Thiophene–Quinoxaline Copolymer}. \emph{Adv. Energy
  Mater.} \textbf{2013}, \emph{3}, 806--814\relax
\mciteBstWouldAddEndPuncttrue
\mciteSetBstMidEndSepPunct{\mcitedefaultmidpunct}
{\mcitedefaultendpunct}{\mcitedefaultseppunct}\relax
\EndOfBibitem
\bibitem[Burke \latin{et~al.}(2015)Burke, Sweetnam, Vandewal, and
  McGehee]{Burke2015}
Burke,~T.~M.; Sweetnam,~S.; Vandewal,~K.; McGehee,~M.~D. {Beyond Langevin
  Recombination: How Equilibrium Between Free Carriers and Charge Transfer
  States Determines the Open-Circuit Voltage of Organic Solar Cells}.
  \emph{Adv. Energy Mater.} \textbf{2015}, \emph{5}, 1500123\relax
\mciteBstWouldAddEndPuncttrue
\mciteSetBstMidEndSepPunct{\mcitedefaultmidpunct}
{\mcitedefaultendpunct}{\mcitedefaultseppunct}\relax
\EndOfBibitem
\bibitem[Shoaee \latin{et~al.}(2019)Shoaee, Armin, Stolterfoht, Hosseini,
  Kurpiers, and Neher]{Shoaee2019}
Shoaee,~S.; Armin,~A.; Stolterfoht,~M.; Hosseini,~S.~M.; Kurpiers,~J.;
  Neher,~D. {Decoding Charge Recombination through Charge Generation in Organic
  Solar Cells}. \emph{Sol. RRL} \textbf{2019}, \emph{3}, 1900184\relax
\mciteBstWouldAddEndPuncttrue
\mciteSetBstMidEndSepPunct{\mcitedefaultmidpunct}
{\mcitedefaultendpunct}{\mcitedefaultseppunct}\relax
\EndOfBibitem
\bibitem[Andersson \latin{et~al.}(2013)Andersson, Melianas, Infahasaeng, Tang,
  Yartsev, Ingan\"{a}s, and Sundstr\"{o}m]{Andersson2013}
Andersson,~L.~M.; Melianas,~A.; Infahasaeng,~Y.; Tang,~Z.; Yartsev,~A.;
  Ingan\"{a}s,~O.; Sundstr\"{o}m,~V. {Unified Study of Recombination in
  Polymer:Fullerene Solar Cells Using Transient Absorption and
  Charge-Extraction Measurements}. \emph{J. Phys. Chem. Lett.} \textbf{2013},
  \emph{4}, 2069--2072\relax
\mciteBstWouldAddEndPuncttrue
\mciteSetBstMidEndSepPunct{\mcitedefaultmidpunct}
{\mcitedefaultendpunct}{\mcitedefaultseppunct}\relax
\EndOfBibitem
\end{mcitethebibliography}


\providecommand{\latin}[1]{#1}
\makeatletter
\providecommand{\doi}
  {\begingroup\let\do\@makeother\dospecials
  \catcode`\{=1 \catcode`\}=2\doi@aux}
\providecommand{\doi@aux}[1]{\endgroup\texttt{#1}}
\makeatother
\providecommand*\mcitethebibliography{\thebibliography}
\csname @ifundefined\endcsname{endmcitethebibliography}
  {\let\endmcitethebibliography\endthebibliography}{}
\begin{mcitethebibliography}{13}
\providecommand*\natexlab[1]{#1}
\providecommand*\mciteSetBstSublistMode[1]{}
\providecommand*\mciteSetBstMaxWidthForm[2]{}
\providecommand*\mciteBstWouldAddEndPuncttrue
  {\def\EndOfBibitem{\unskip.}}
\providecommand*\mciteBstWouldAddEndPunctfalse
  {\let\EndOfBibitem\relax}
\providecommand*\mciteSetBstMidEndSepPunct[3]{}
\providecommand*\mciteSetBstSublistLabelBeginEnd[3]{}
\providecommand*\EndOfBibitem{}
\mciteSetBstSublistMode{f}
\mciteSetBstMaxWidthForm{subitem}{(\alph{mcitesubitemcount})}
\mciteSetBstSublistLabelBeginEnd
  {\mcitemaxwidthsubitemform\space}
  {\relax}
  {\relax}

\bibitem[Howard \latin{et~al.}(2014)Howard, Etzold, Laquai, and
  Kemerink]{Howard2014}
Howard,~I.~A.; Etzold,~F.; Laquai,~F.; Kemerink,~M. {Nonequilibrium Charge
  Dynamics in Organic Solar Cells}. \emph{Adv. Energy Mater.} \textbf{2014},
  \emph{4}, 1301743\relax
\mciteBstWouldAddEndPuncttrue
\mciteSetBstMidEndSepPunct{\mcitedefaultmidpunct}
{\mcitedefaultendpunct}{\mcitedefaultseppunct}\relax
\EndOfBibitem
\bibitem[Melianas \latin{et~al.}(2014)Melianas, Pranculis, Devi\v{z}is,
  Gulbinas, Ingan\"{a}s, and Kemerink]{Melianas2014}
Melianas,~A.; Pranculis,~V.; Devi\v{z}is,~A.; Gulbinas,~V.; Ingan\"{a}s,~O.;
  Kemerink,~M. {Dispersion-Dominated Photocurrent in Polymer:Fullerene Solar
  Cells}. \emph{Adv. Funct. Mater.} \textbf{2014}, \emph{24}, 4507--4514\relax
\mciteBstWouldAddEndPuncttrue
\mciteSetBstMidEndSepPunct{\mcitedefaultmidpunct}
{\mcitedefaultendpunct}{\mcitedefaultseppunct}\relax
\EndOfBibitem
\bibitem[Melianas \latin{et~al.}(2015)Melianas, Etzold, Savenije, Laquai,
  Ingan\"{a}s, and Kemerink]{Melianas2015}
Melianas,~A.; Etzold,~F.; Savenije,~T.~J.; Laquai,~F.; Ingan\"{a}s,~O.;
  Kemerink,~M. {Photo-Generated Carriers Lose Energy During Extraction from
  Polymer-Fullerene Solar Cells}. \emph{Nat. Commun.} \textbf{2015}, \emph{6},
  8778\relax
\mciteBstWouldAddEndPuncttrue
\mciteSetBstMidEndSepPunct{\mcitedefaultmidpunct}
{\mcitedefaultendpunct}{\mcitedefaultseppunct}\relax
\EndOfBibitem
\bibitem[Melianas \latin{et~al.}(2017)Melianas, Pranculis, Xia, Felekidis,
  Ingan\"{a}s, Gulbinas, and Kemerink]{Melianas2017}
Melianas,~A.; Pranculis,~V.; Xia,~Y.; Felekidis,~N.; Ingan\"{a}s,~O.;
  Gulbinas,~V.; Kemerink,~M. {Photogenerated Carrier Mobility Significantly
  Exceeds Injected Carrier Mobility in Organic Solar Cells}. \emph{Adv. Energy
  Mater.} \textbf{2017}, \emph{7}, 1602143\relax
\mciteBstWouldAddEndPuncttrue
\mciteSetBstMidEndSepPunct{\mcitedefaultmidpunct}
{\mcitedefaultendpunct}{\mcitedefaultseppunct}\relax
\EndOfBibitem
\bibitem[Felekidis \latin{et~al.}(2018)Felekidis, Melianas, and
  Kemerink]{Felekidis2018}
Felekidis,~N.; Melianas,~A.; Kemerink,~M. Automated Open-Source Software for
  Charge Transport Analysis in Single-Carrier Organic Semiconductor Diodes.
  \emph{Org. Electron.} \textbf{2018}, \emph{61}, 318--328\relax
\mciteBstWouldAddEndPuncttrue
\mciteSetBstMidEndSepPunct{\mcitedefaultmidpunct}
{\mcitedefaultendpunct}{\mcitedefaultseppunct}\relax
\EndOfBibitem
\bibitem[Murthy \latin{et~al.}(2013)Murthy, Melianas, Tang, Ju\v{s}ka,
  Arlauskas, Zhang, Siebbeles, Ingan\"as, and Savenije]{Murthy2013}
Murthy,~D. H.~K.; Melianas,~A.; Tang,~Z.; Ju\v{s}ka,~G.; Arlauskas,~K.;
  Zhang,~F.; Siebbeles,~L. D.~A.; Ingan\"as,~O.; Savenije,~T.~J. Origin of
  Reduced Bimolecular Recombination in Blends of Conjugated Polymers and
  Fullerenes. \emph{Adv. Funct. Mater.} \textbf{2013}, \emph{23},
  4262--4268\relax
\mciteBstWouldAddEndPuncttrue
\mciteSetBstMidEndSepPunct{\mcitedefaultmidpunct}
{\mcitedefaultendpunct}{\mcitedefaultseppunct}\relax
\EndOfBibitem
\bibitem[Tvingstedt \latin{et~al.}(2009)Tvingstedt, Vandewal, Gadisa, Zhang,
  Manca, and Ingan\"{a}s]{Tvingstedt2009}
Tvingstedt,~K.; Vandewal,~K.; Gadisa,~A.; Zhang,~F.; Manca,~J.; Ingan\"{a}s,~O.
  {Electroluminescence from Charge Transfer States in Polymer Solar Cells}.
  \emph{J. Am. Chem. Soc.} \textbf{2009}, \emph{131}, 11819--11824\relax
\mciteBstWouldAddEndPuncttrue
\mciteSetBstMidEndSepPunct{\mcitedefaultmidpunct}
{\mcitedefaultendpunct}{\mcitedefaultseppunct}\relax
\EndOfBibitem
\bibitem[Heiber \latin{et~al.}(2015)Heiber, Baumbach, Dyakonov, and
  Deibel]{Heiber2015}
Heiber,~M.~C.; Baumbach,~C.; Dyakonov,~V.; Deibel,~C. Encounter-Limited
  Charge-Carrier Recombination in Phase-Separated Organic Semiconductor Blends.
  \emph{Phys. Rev. Lett.} \textbf{2015}, \emph{114}, 136602\relax
\mciteBstWouldAddEndPuncttrue
\mciteSetBstMidEndSepPunct{\mcitedefaultmidpunct}
{\mcitedefaultendpunct}{\mcitedefaultseppunct}\relax
\EndOfBibitem
\bibitem[Burke \latin{et~al.}(2015)Burke, Sweetnam, Vandewal, and
  McGehee]{Burke2015}
Burke,~T.~M.; Sweetnam,~S.; Vandewal,~K.; McGehee,~M.~D. {Beyond Langevin
  Recombination: How Equilibrium Between Free Carriers and Charge Transfer
  States Determines the Open-Circuit Voltage of Organic Solar Cells}.
  \emph{Adv. Energy Mater.} \textbf{2015}, \emph{5}, 1500123\relax
\mciteBstWouldAddEndPuncttrue
\mciteSetBstMidEndSepPunct{\mcitedefaultmidpunct}
{\mcitedefaultendpunct}{\mcitedefaultseppunct}\relax
\EndOfBibitem
\bibitem[Loos \latin{et~al.}(2009)Loos, Sourty, Lu, {de With}, and {van
  Bavel}]{Loos2009}
Loos,~J.; Sourty,~E.; Lu,~K.; {de With},~G.; {van Bavel},~S. Imaging Polymer
  Systems with High-Angle Annular Dark Field Scanning Transmission Electron
  Microscopy (HAADF-STEM). \emph{Macromol.} \textbf{2009}, \emph{42},
  2581--2586\relax
\mciteBstWouldAddEndPuncttrue
\mciteSetBstMidEndSepPunct{\mcitedefaultmidpunct}
{\mcitedefaultendpunct}{\mcitedefaultseppunct}\relax
\EndOfBibitem
\bibitem[Alekseev \latin{et~al.}(2015)Alekseev, Hedley, Al-Afeef, Ageev, and
  Samuel]{Alekseev2015}
Alekseev,~A.; Hedley,~G.~J.; Al-Afeef,~A.; Ageev,~O.~A.; Samuel,~I. D.~W.
  {Morphology and Local Electrical Properties of PTB7:PC71BM Blends}. \emph{J.
  Mater. Chem. A} \textbf{2015}, \emph{3}, 8706--8714\relax
\mciteBstWouldAddEndPuncttrue
\mciteSetBstMidEndSepPunct{\mcitedefaultmidpunct}
{\mcitedefaultendpunct}{\mcitedefaultseppunct}\relax
\EndOfBibitem
\bibitem[B\"acke \latin{et~al.}(2015)B\"acke, Lindqvist, {de Zerio Mendaza},
  Gustafsson, Wang, Andersson, M\"uller, and Olsson]{Backe2015}
B\"acke,~O.; Lindqvist,~C.; {de Zerio Mendaza},~A.~D.; Gustafsson,~S.;
  Wang,~E.; Andersson,~M.~R.; M\"uller,~C.; Olsson,~E. {Mapping Fullerene
  Crystallization in a Photovoltaic Blend: An Electron Tomography Study}.
  \emph{Nanoscale} \textbf{2015}, \emph{7}, 8451--8456\relax
\mciteBstWouldAddEndPuncttrue
\mciteSetBstMidEndSepPunct{\mcitedefaultmidpunct}
{\mcitedefaultendpunct}{\mcitedefaultseppunct}\relax
\EndOfBibitem
\end{mcitethebibliography}

\end{document}


\newpage
\paragraph{Discussion of Symmetrized Hopping Parameters} A well-known problem of kinetic Monte Carlo~(KMC) simulations of hopping transport in energetically disordered media is that calculation times explode when the normalized disorder~$\hat\sigma = \sigma/kT$ becomes larger than 3--4. An additional problem arises in bipolar systems if the attempt to hop frequencies~$\nu_0$ of the two charge species~(electrons and holes) differ by more than roughly an order of magnitude. In this case, the hopping events of the species with the highest~$\nu_0$ outnumber those of the slower species. This leads to poor statistics for the slower species that can only be cured by increasing the total number of hops considered in the simulation, i.e. by increasing the total calculation time. Unfortunately, the previously determined hopping parameters for the TQ1:PCBM system studied here cause both problems to arise.

For holes in TQ1 we found in Refs.~\citenum{Howard2014,Melianas2014,Melianas2015,Melianas2017,Felekidis2018} $\sigma_\text{h} \approx \unit[50\text{--}100]{meV}$ and $\nu_{0,\text{h}} \approx \unit[0.1\text{--}1 \times 10^{10}]{s^{-1}}$ and for electrons in PCBM $\sigma_\text{e} \approx \unit[120]{meV}$ and $\nu_{0,\text{e}} \approx \unit[1 \times 10^{13}]{s^{-1}}$. In order to keep calculation times to acceptable levels~(days), we had to use partially symmetrized hopping and disorder parameters. In this, we could make use of the previously observed interchangeability of the disorder and the attempt-to-hop frequency,\cite{Melianas2014} where increases in one of the two parameters can be largely compensated by a simultaneous increase in the other parameter. Bearing the above in mind, we used~$\sigma_\text{e} = \sigma_\text{h} = \unit[75]{meV}$ for both electrons and holes, and $\nu_{0,\text{e}} = \unit[1 \times 10^{11}]{s^{-1}}$ and $\nu_{0,\text{h}} = \unit[1 \times 10^{10}]{s^{-1}}$, keeping the corresponding steady state electron mobility significantly larger than that of the holes. The other parameter that has (a minor) influence on the hopping process in the used model is the nearest neighbor distance $a_\text{NN} = \unit[1.8]{nm}$ that we fixed to the value that we obtained from temperature-dependent charge transport studies.\cite{Felekidis2018} A consequence of the symmetrized hopping parameters is that the other rates in the simulation, specifically the recombination rate of the CT exciton, become relative to these values. This explains why the used CT rate $\kct = \unit[3 \times 10^7]{s^{-1}}$ is two to three times lower than the typical values used before.\cite{Howard2014,Melianas2015}

\newpage
\paragraph{Charge Recombination in TQ1:P\ce{C71}BM} The three relevant transitions for the recombination of photogenerated charges are the separation rate of CT~states~($\kd$), the encounter rate of free electrons and holes~($\kenc$), and the decay rate of CT~states into the ground state~($\kct$). Importantly, the encounter complex formed by two independent carriers has been identified as the same CT~state as involved in charge separation.\cite{Murthy2013,Tvingstedt2009} Only if the rate at which CT~states recombine were much faster than the rate at which they dissociate~($\kct \gg \kd$), the recombination would be encounter-limited and Langevin theory applicable. In practice, however, virtually all OPV~blends show recombination rates that are substantially reduced compared to the Langevin model. The apparent steady-state recombination rate constant of~$\unit[6 \times 10^{-18}]{m^3/s}$ we assume Section~2.3 in the main text for TQ1:P\ce{C71}BM implies a reduction of 2~orders of magnitude.

Such large reduction factors cannot be explained by geometrical confinement, i.e., the fact that electrons and holes are attributed to different material phases and only meet at the heterointerface.\cite{Heiber2015} We must therefore assume that recombination in TQ1:P\ce{C71}BM is not encounter-limited. This is reasonable considering the following: although the details of charge separation are not yet well understood, the rate~$\kd$ will depend on the ability of the two involved carriers to move away from each other. Still being a hopping process, this will happen with a rate on the order of $10^{10}$ to $\unit[10^{13}]{s^{-1}}$. Because this is much faster than the rate at which the CT~state decays into the ground state, i.e., its inverse lifetime~($\kct = 10^7$ to $\unit[10^8]{s^{-1}}$), there is enough time to establish an equilibrium between CT~states and free electrons and holes.\cite{Burke2015,Howard2014} In other words, the carriers forming the CT~state have multiple attempts to escape from their mutual Coulomb attraction before they ultimately recombine. One can easily see that it is then the rate~$\kct$ which determines the position of the equilibrium; decreasing~$\kct$ will shift the equilibrium more towards dissociation, effectively decreasing the charge recombination rate detectable experimentally.

\newpage
\begin{figure*}[h]
\includegraphics[width=0.5\linewidth]{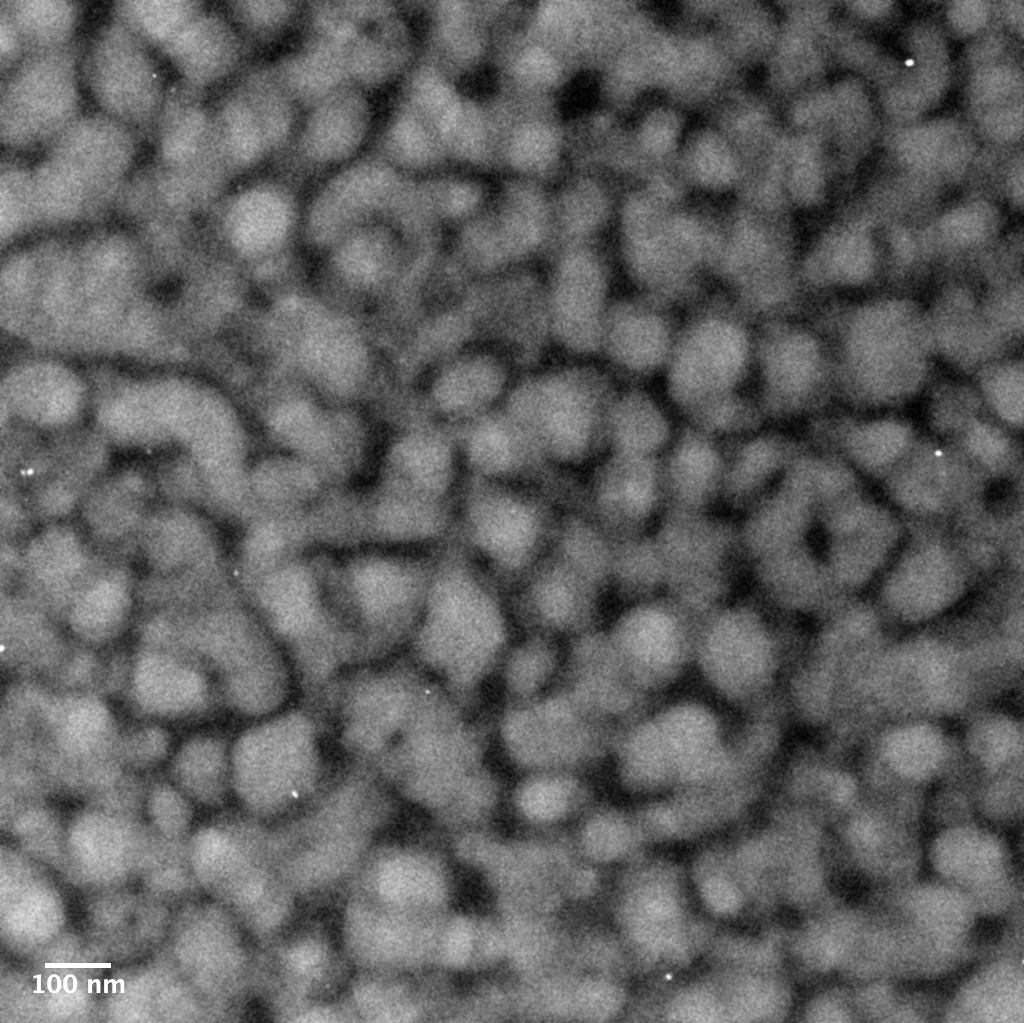}
\caption{Transmission electron micrograph in HAADF-STEM mode of the same TQ1:P\ce{C71}BM blend film as discussed in the main text. The HAADF signal in STEM originates from electrons that have inelastically scattered to high angles when transmitted through the sample. The signal intensity is higher if the electrons scatter against molecules of higher average atomic number~(Z-number). This is likely not relevant here, since the TQ1 monomer and the P\ce{C71}BM molecule have a similar average Z-number. The signal intensity is also increased for regions of higher density or thickness. Since the film thickness is rather uniform, the bright regions in the HAADF-STEM image show areas of higher density. This correlates well with what is known about fullerenes and their aggregates.\cite{Loos2009,Alekseev2015,Backe2015}.}
\end{figure*}

\newpage
\begin{figure*}[h]
\includegraphics[width=0.7\linewidth]{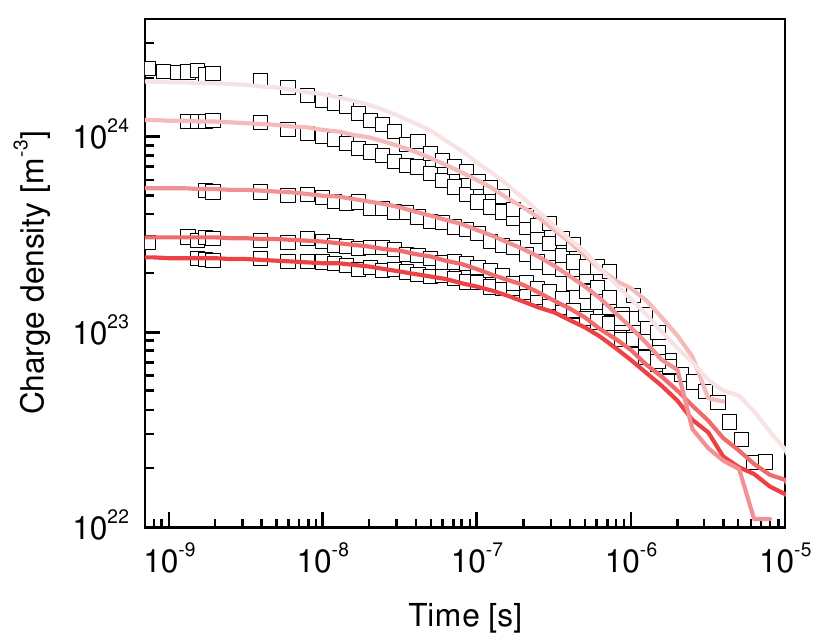}
\caption{KMC simulations of transient absorption data for non-symmetrized hopping parameters. For the electrons $\sigma_\text{e} = \unit[120]{meV}$ and $\nu_{0,\text{e}} = \unit[1 \times 10^{13}]{s^{-1}}$ was assumed and for the holes $\sigma_\text{h} = \unit[83]{meV}$ and $\nu_{0,\text{h}} = \unit[5 \times 10^{9}]{s^{-1}}$, according to previous experimental work.\cite{Melianas2015} As mentioned in the text above, using these parameters required us to assume a slightly higher CT rate of $\kct = \unit[8 \times 10^7]{s^{-1}}$.}
\end{figure*}

\bibliography{supplement}